\documentclass[letterpaper]{article} 
\usepackage[]{aaai25}  
\usepackage{times}  
\usepackage{helvet}  
\usepackage{courier}  
\usepackage[hyphens]{url}  
\usepackage{graphicx} 
\usepackage{natbib}  
\usepackage{caption} 
\usepackage{algorithm}
\usepackage{algorithmic}

\usepackage{newfloat}
\usepackage{listings}
\DeclareCaptionStyle{ruled}{labelfont=normalfont,labelsep=colon,strut=off} 
\floatstyle{ruled}
\newfloat{listing}{tb}{lst}{}
\floatname{listing}{Listing}
\title{Mechanistic Understanding of Language Models in Syntactic Code Completion}
\author{
    Samuel Miller\equalcontrib,\:
    Daking Rai\equalcontrib,\:
    Ziyu Yao
}
\affiliations{
    George Mason University \\
     smille20@gmu.edu, 
     drai2@gmu.edu,	
     ziyuyao@gmu.edu
}
\usepackage[dvipsnames]{xcolor}

\usepackage{booktabs}
\usepackage{amsmath}

\usepackage{listings}
\usepackage{graphicx}
\usepackage{subcaption} 
\lstdefinestyle{python}{
    language=Python,
    basicstyle=\ttfamily\footnotesize,
    keywordstyle=\color{blue}\bfseries,
    stringstyle=\color{green!50!black},
    commentstyle=\color{gray}\itshape,
    showstringspaces=false,
    frame=single,
    numbers=left,
    numberstyle=\tiny\color{gray},
    backgroundcolor=\color{gray!10}
}
\begin{document}

\maketitle

\begin{abstract}
{Recently, language models (LMs) have shown impressive proficiency in code generation tasks, especially when fine-tuned on code-specific datasets, commonly known as Code LMs. However, our understanding of the internal decision-making processes of Code LMs, such as how they use their (syntactic or semantic) knowledge, remains limited, which could lead to unintended harm as they are increasingly used in real life. This motivates us to conduct one of the first Mechanistic Interpretability works to understand how Code LMs perform a syntactic completion task, specifically the closing parenthesis task, on the CodeLlama-7b model~\cite{roziere2023code}. Our findings reveal that the model requires middle-later layers until it can confidently predict the correct label for the closing parenthesis task. Additionally, we identify that while both multi-head attention (MHA) and feed-forward (FF) sub-layers play essential roles, MHA is particularly crucial. Furthermore, we also discover attention heads that keep track of the number of already closed parentheses precisely but may or may not promote a correct number of closing parentheses that are still missing, leading to a positive or negative impact on the model's performance.
} 
 
\end{abstract}

\section{Introduction}

{Language models (LMs) have recently showcased impressive capabilities in code-related tasks, generating executable code from task specifications provided in natural language prompts~\cite{jiang2024survey, zan2023large}. To further enhance their capability, they are often instruction-tuned on code-specific datasets, resulting in specialized models known as Code LMs, such as CodeLlama~\cite{roziere2023code}, StarCoder~\cite{li2023starcoder}, Code Gemma~\cite{team2024codegemma}, and DeepSeek-Coder~\cite{guo2024deepseek}. These advancements have led to widespread adoption by programmers and researchers, integrating LMs into their daily workflows to assist with coding tasks~\cite{dakhel2023github}.}  

{However, despite substantial progress in code generation, Code LMs can generate incorrect code, particularly when handling complex tasks~\cite{dou2024s, tambon2024bugs}. These faulty codes can pose significant risks, especially when used by novice programmers working on critical applications~\cite{dakhel2023github}. Furthermore, even when the generated code snippets are functionally correct, recent research found that these code snippets could contain security vulnerabilities~\cite{siddiq2022securityeval, yang2024seccodeplt}. Understanding the capabilities and limitations of Code LMs, such as where and how they invoke relevant knowledge internally, is therefore essential to mitigate potential harm in high-stakes scenarios. Yet, there is limited knowledge about the internals of these LMs while generating code for a given task. Developing deeper insights into their knowledge-relevant mechanisms is crucial for improving their reliability, performance, and safe deployment.}

Mechanistic interpretability (MI) has recently emerged as a promising approach to understanding the internal mechanisms of LMs~\cite{olah2020zoom, elhage2021mathematical, rai2024investigation, bereska2024mechanistic}. MI studies have investigated a range of LM behaviors, including in-context learning~\cite{elhage2021mathematical, bansal2022rethinking, ren2024identifying}, reasoning~\cite{stolfo2023mechanistic, rai2024investigation, dutta2024think}, and fact recall~\cite{geva2023dissecting, chughtai2024summing}, providing valuable insights into how various LM components, such as multi-head attention (MHA) and feedforward (FF) sublayers, contribute to these capabilities. While substantial work has been done to investigate various behaviors of LMs, there has been limited focus on understanding how Code LMs internally use their knowledge in code generation tasks.

To address this gap, we present one of the first MI studies on Code LMs, where we investigate the internal workings of the CodeLlama-7b~\cite{roziere2023code} for the syntax completion task, the success of which requires an LM to not only locate its declarative knowledge of the programming language but also use the knowledge with its other capabilities (e.g., counting) smartly.
Specifically, we study how CodeLlama-7b performs the closing parentheses task (e.g., \lstinline|print(str(1| $\rightarrow$ \lstinline|))|), where each opening parenthesis must be paired with a closing parenthesis. To this end, we first contribute a synthetic dataset for systematically studying a Code LM's syntax completion performance. Our dataset includes a total 168 prompts covering three sub-tasks with the number of closing parentheses in the target tokens being 2, 3, and 4, respectively. These prompts include recursive calls of class constructors including \lstinline|str|, \lstinline|list|, and \lstinline|set|, with the number of open parentheses ranging from 2 to 12. With our dataset,
we perform a series of analyses investigating the internal mechanisms of CodeLlama-7b, including projecting the intermediate (sub-)layers' activations via the logit lens~\cite{nostalgebraist2020blog} to understand the typical timing when the model realizes the correct token, measuring the logit difference between the correct and the counterfactual tokens to understand the effective contribution of each (sub-)layer to correct token predictions, and performing attention visualization analysis to discover the attention patterns inside the Code LM. 
Our experimental results reveal that: 
\begin{itemize}
    \item The Code LM can realize the correct target token only from the middle-to-late layers. For example, in case of needing two closing parentheses in the target token, the model on average only ranks the correct token within top 10 based on its projected logit value from layer 18, and being the top 1 from layer 25. We also discover that when the required closing parentheses increase, it becomes even more difficult for the model to identify the correct token, as illustrated by its even later layers for ranking the correct token within the top 10 or top 1.
    \item When looking into a comparative effect of predicting the correct token vs. predicting the counterfactual token, both MHA and FF sub-layers contribute to the task. However, MHA sub-layers make a more critical contribution to the prediction of the correct tokens. In addition, the contributions of the (sub-)layers generally follow a similar pattern when we vary the number of open parentheses and the class constructors in the input prompt.
    \item Finally, we identified and interpreted key attention heads responsible for performing the task. For instance, we discovered two attention heads, $L30H0$ and $L27H24$, both keeping track of the number of already closed parentheses precisely.
    However, while $L30H0$ consistently promotes the correct number of closing parentheses that are still missing across sub-tasks, $L27H24$ always promotes the token including exactly two closing parentheses, which we summarize as \emph{incorrect knowledge association}. As a consequence, it was found to be crucial when the task requires two closing parentheses, but has a negative effect otherwise.
\end{itemize}


\section{Methodology and Dataset}\label{sec:dataset-and-methodology}

In this section, we introduce our methodology towards forming a mechanistic understanding of how a Code LM performs a syntax completion task that requires using its knowledge about a programming language in combination with others (e.g., counting). Our experiment will focus on Python code generation using CodeLlama-7b~\cite{roziere2023code}, which is a state-of-the-art (SOTA) medium-size Code LM with a 32-layer decoder-only transformer architecture. The specific model checkpoint we use is ``CodeLlama-7b-hf'' (i.e., the base model). In what follows, we will first give an overview of our experiment design, then present our process of dataset generation to facilitate the experiment, and finally describe the methods we will use to analyze a Code LM.

\begin{table*}[t!]
    \centering
    \resizebox{\textwidth}{!}{
    \begin{tabular}{cclcc}
         \toprule
         \textbf{Sub-Task} & \textbf{\#of Examples} & \textbf{Prompt $\rightarrow$ Target Token} & \textbf{Counterfactual Token} & \textbf{Accuracy}   \\
         \midrule 
          Two Closing Paren & 56 &  {\#print a list containing 2\textbackslash n \lstinline|print(list(list(tuple([2]))| $\rightarrow$ \lstinline|))|} & \lstinline|)| & 100.0\% \\
         \midrule
         Three Closing Paren & 84 &  \#print a string 12\textbackslash n\lstinline|print(str(str(12| $\rightarrow$ \lstinline|)))| & \lstinline|))| & 76.2\%\\
         \midrule
         Four Closing Paren & 28 &  \#print a set containing 123\textbackslash n\lstinline|print(set(set(set(set(tuple([123]))| $\rightarrow$ \lstinline|))))| & \lstinline|)))| & 100.0\% \\
         \bottomrule
    \end{tabular}
    }
    \caption{Examples of prompts provided to the Code LM for each sub-task. In our work, we synthesize a dataset of 168 prompts covering three sub-tasks with the number of closing parentheses in the target tokens being 2, 3, and 4, respectively. The sub-task design is based on the specific Code LM (Codellama)'s tokenization effect (see Section~\ref{subsec:data-generation}).}
   
    \label{tab:subtask_prompts}
\end{table*}

\subsection{Motivation and Overview}\label{subsec:overview-experiment-design}

Syntax completion is a crucial and fundamental part of LM code generation, as syntactic correctness is essential for code to be executable. \citet{dou2024s} recently found that even SOTA Code LMs still suffer from syntactic issues to various extents in their generation. This has motivated us to carefully understand how a Code LM performs syntax completion. In our work, we select the closing parentheses task (e.g., \lstinline|print(str(1| $\rightarrow$ \lstinline|))|) as our task, given it is one of the most common syntactical structures seen across programming languages; as a result, it is safe to assume that SOTA Code LMs have learned the necessary syntactic knowledge from their training. The input provided to a Code LM in this task is a partially complete line of code {(e.g., {\lstinline|print(str(1|})}, which includes a varying number of function or class constructor calls but is missing some \emph{final} number of closing parentheses that needs to be predicted as a whole in its next token.\footnote{Modern Code LMs may or may not predict parentheses one by one. For example, Codellama tokenizes \lstinline|print(str(1))| into \lstinline|print|, \lstinline|(|, \lstinline|str|, \lstinline|(|, \lstinline|1|, and \lstinline|))|, a total of 5 tokens, with the two closing parentheses being predicted as one single token. In practice, we have observed that when Codellama completes the syntax correctly, it all follows this tokenization practice, namely, it directly generates ``\lstinline|))|'' as a single token instead of generating ``\lstinline|)|'' for two times. Therefore, in our task design, we consistently use the last token based on the Code LM's tokenization as the target token when analyzing the model's syntax completion performance.} 
The Code LM is then tasked with predicting the next token that consists of the necessary number of closing parentheses for the line of code to be syntactically correct {(e.g., {\lstinline|))| in the running example})}. 

In our experiments, we focus on analyzing the Code LM in Python programming language.
This task was further broken down into three sub-tasks based on the number of closing parentheses required to correctly complete the line of code, which were two, three, and four closing parentheses. A partially completed line of code example for each of the sub-tasks can be seen in Table~\ref{tab:subtask_prompts}.

\subsection{Data Generation} \label{subsec:data-generation}
To evaluate the Code LM on the closing parentheses task, we created an initial synthetic dataset consisting of 168 input prompts, with each prompt consisting of a simple natural language instruction, in the form of a code comment that describes the desired semantic meaning of the following line of code, and a partially completed line of Python code. In our preliminary exploration, we found the code comment to be necessary to avoid semantic ambiguity, as otherwise there could be infinite plausible continuations of the same line of code (e.g., continuing ``\lstinline|print(str(1|'' with more digits), which will make the analysis difficult.
We began the dataset preparation process by searching for Python functions that were both commonly used in practice and could accept arguments of varying data types. To this end, we decided to initially focus on generating prompts that utilized the built-in {\lstinline|print|} function while varying the argument supplied to the function. The argument was varied through the selection of a Python built-in class constructor, from a set containing {\lstinline|str|}, {\lstinline|list|}, and {\lstinline|set|}, the integer value passed to the constructor, and the number of nested constructor calls (with the number of open parentheses ranging from 2 to 12 in our data synthesis), which was used to vary the number of required closing parentheses. Following, the various generated arguments were combined with a {\lstinline|print|} function call to produce completed lines of code. The completed lines of code were then combined with their respective natural language instructions and tokenized using the Code LM (i.e., Codellama)'s tokenizer to produce the final prompts along with their associated correct next (and final) token. 

On this dataset, Codellama-7b was able to achieve an overall accuracy of 88\%, with the breakdown for each sub-task shown in Table~\ref{tab:subtask_prompts}. 
Intriguingly, we observed that Codellama-7b achieves a lower accuracy on the Three Closing Parentheses sub-task than on the Four Closing Parentheses sub-task. Upon examination, we found that all failing cases in the Three Closing Parentheses sub-task occurred when the number of open parentheses in the prompt ranged from nine to eleven, whereas for the prompts in the Four Closing Parentheses sub-task, the largest number of open parentheses is only eight.\footnote{We note that the range of the number of open parentheses in each sub-task is largely an effect of Codellama's tokenization and cannot be enforced during the dataset generation.} An intuitive conjecture is that, when the number of open parentheses gets larger, it becomes more challenging for the Code LM to correctly count both the number of open parentheses and the existing number of closing parentheses in the prompt, as well as calculate the difference as the required number of closing parentheses for next token prediction.

\subsection{Methodology} \label{subsec:methodology}
We will understand how a Code LM completes a syntax completion task by mainly looking at the model's behaviors of predicting the correct next token, which requires proper use of its knowledge about the programming language. Specifically, we employ \emph{logit lens} or \emph{direct logit attribution}~\cite{nostalgebraist2020blog} to analyze the contribution of each layer and its sublayers (MHA and FF) in predicting the correct next token. \emph{Logit lens} allows us to view what the LM would have predicted in a given (sub-)layer by projecting the intermediate activations (denoted as $v \in {R}^d$, where $d$ is the LM dimension) onto the logit distribution through multiplication with the unembedding parameter matrix (denoted as $W_U \in {R}^{d \times |\mathcal{V}|}$, where $\mathcal{V}$ is the vocabulary set and $|\mathcal{V}|$ denotes its size), i.e., $v W_U \in \mathcal{R}^{|\mathcal{V}|}$. As a result, we can examine the top-$k$ candidate tokens for the next token prediction at each intermediate (sub-)layer by viewing the logit distribution. We refer readers to \citet{nostalgebraist2020blog} or the recent survey paper of \citet{rai2024practical} for a systematic and detailed explanation of the logit lens method.

In addition to analyzing the absolute logit value of the target token, our experiment also involves calculating the \emph{logit difference} to evaluate the contribution of (sub-)layer for the correct token (e.g. {``\lstinline|)))|''}) relative to a \emph{counterfactual} token representing \emph{incorrect knowledge} (e.g. {``\lstinline|))|''}), effectively filtering out (sub-)layers that indiscriminately increase the logit values of several tokens. We list the counterfactual tokens in Table~\ref{tab:subtask_prompts}. Such a comparative analysis has been widely adopted by prior work \cite{vig2020investigating, meng2022locating, wang2022interpretabilitywildcircuitindirect} for effectively discovering task-specific LM behaviors. When the logit difference becomes more positive (or negative, respectively), it implies that the LM is more (or less, respectively) capable of distinguishing between the correct and the misleading tokens.

Finally, when we are able to locate attention layers and heads making the most contribution to the prediction of the correct token, we will then apply \emph{attention visualization} to scrutinize the model's attention pattern. In our experiment, we have found it a very helpful approach for interpreting the LM's behaviors in a human-understandable way.

{Our experiments were carried out using the TransformerLens library~\cite{nanda2022transformerlens} to implement the logit lens and logit difference, and CircuitsVis~\cite{cooney2023circuitsvis} for attention visualization.}



\section{Experimental Results}

\subsection{Overview of Experiments}
Our experiments aim to answer three Research Questions (RQs). RQ1 and RQ2 examine the layer-wise phenomena in the process of the Code LLM generating the correct next token. Specifically, RQ1 leverages \textit{logit lens} to understand what the model would have predicted in a given layer, from which we gauge from which layer the model typically can start picking the correct token. RQ2 then looks into how (e.g., effect of promotion~\cite{geva2022transformer}) each (sub-)layer contributes to the prediction of the correct token, particularly by contrasting the logit values between the correct token and a \emph{counterfactual} token (Table~\ref{tab:subtask_prompts}). In this process, we also locate the critical attention layers that strongly promote the generation of the correct token. Following that, RQ3 then specifically investigates the patterns of these critical attention layers, such as how different attention heads in these layers play a role and how each attention head functions, aiming to form a clearer understanding of how the Code LM becomes aware of the required number of closing parentheses.

\begin{table*}[t!]
    \centering
    \resizebox{\textwidth}{!}{
    \begin{tabular}{cccc}
         \toprule
         \textbf{Sub-Task} & \textbf{$L_{\text{Top10}}$} & \textbf{$L_{\text{Top1}}$} & \textbf{$L_{\text{ConsistentTop1}}$}   \\
         \midrule 
          Two Closing Paren & 
            \includegraphics[width=0.3\textwidth]{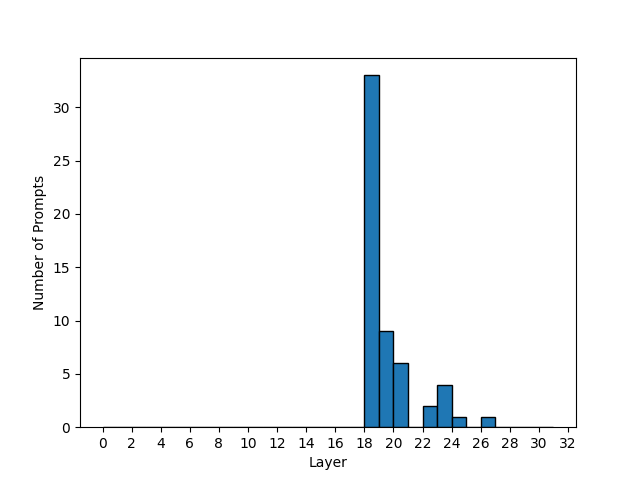} &  
            \includegraphics[width=0.3\textwidth]{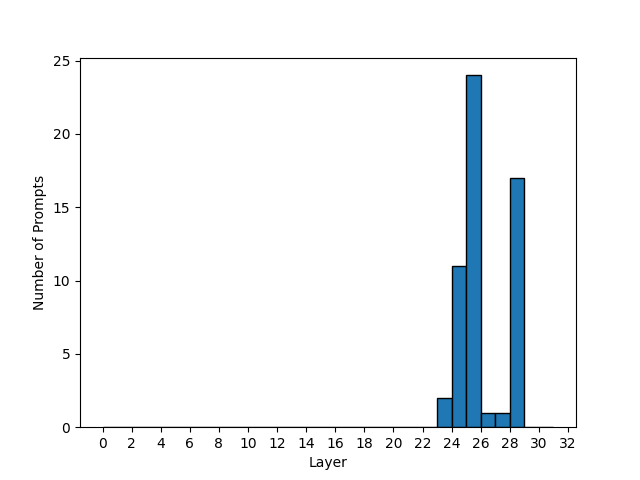} & 
            \includegraphics[width=0.3\textwidth]{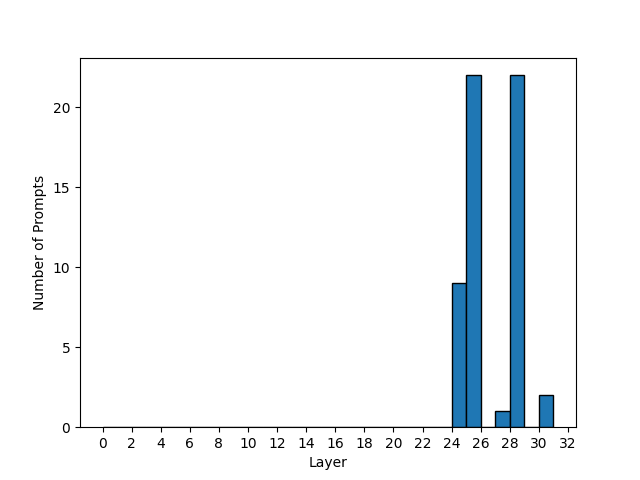}
        \\
         \midrule
         Three Closing Paren & 
            \includegraphics[width=0.3\textwidth]{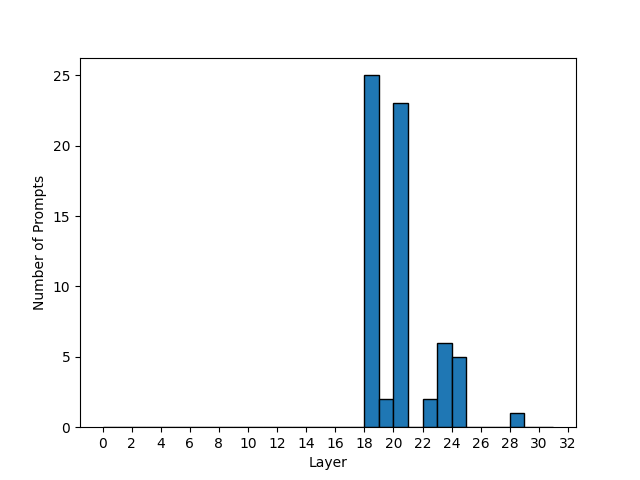} &  
            \includegraphics[width=0.3\textwidth]{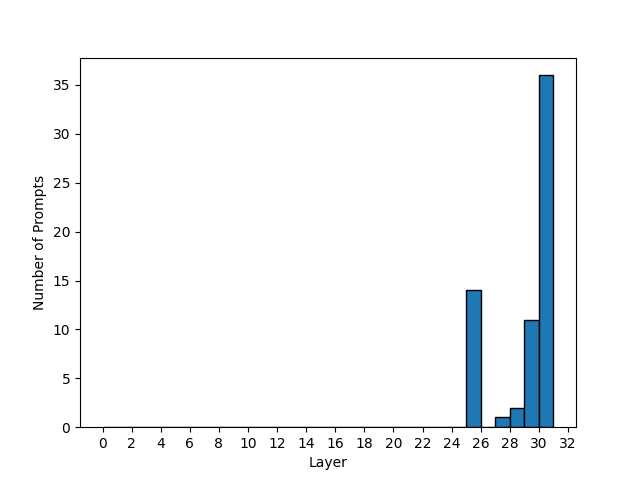} & 
            \includegraphics[width=0.3\textwidth]{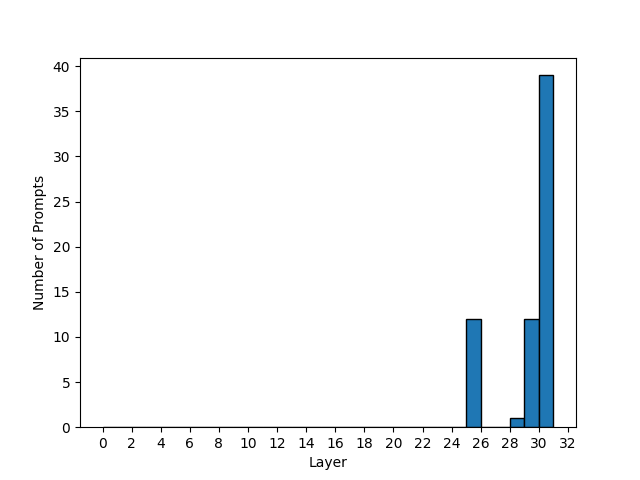}
        \\
         \midrule
         Four Closing Paren & 
            \includegraphics[width=0.3\textwidth]{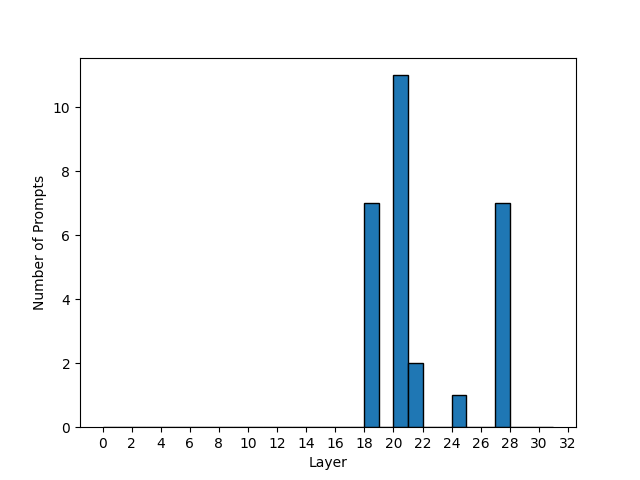} &  
            \includegraphics[width=0.3\textwidth]{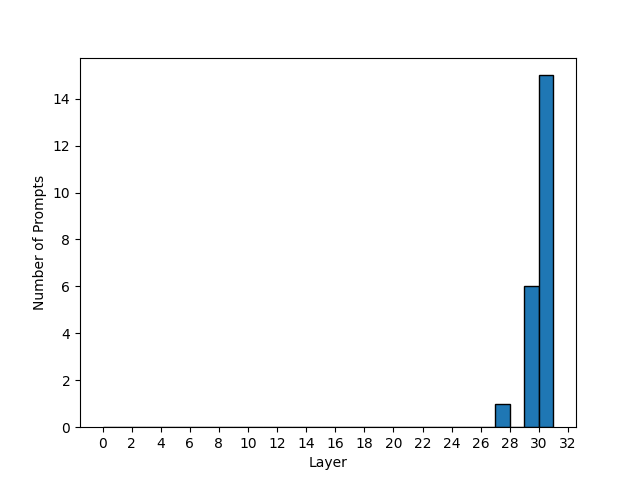} & 
            \includegraphics[width=0.3\textwidth]{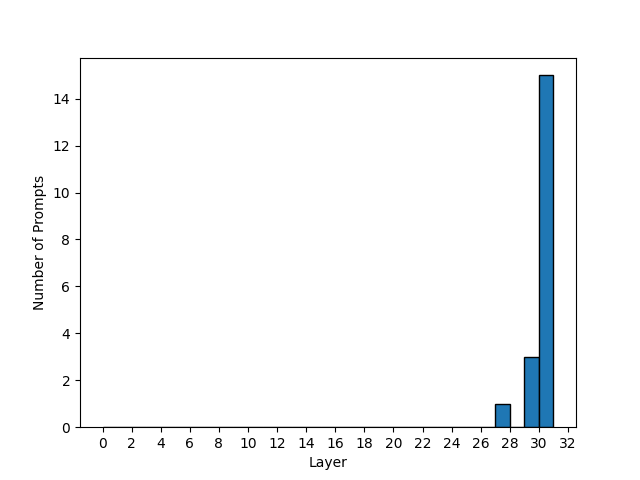}
        \\
         \bottomrule
    \end{tabular}
    }
    \caption{{Aggregated at the sub-task level, we report the layer distribution when the correct token's logit value is ranked within top 10 ($L_{\text{Top10}}$), top 1 ($L_{\text{Top1}}$), and consistently top 1 afterward ($L_{\text{ConsistentTop1}}$), respectively.
    We observed that, compared to the Three and the Four Closing Parenthesis sub-tasks, on the Two Closing Parenthesis sub-task the model can identify the correct token at an earlier layer, implying that the latter sub-task is considered easier than the former two.
    }}
    \label{tab:subtask_top_ten_logits}
\end{table*}

\subsection{RQ1: At what layer does the Code LM start picking the correct token?}\label{subsec:exp1}
The overarching goal of this RQ is to better understand at what points during an inference phase the Code LM has an understanding of what the next token prediction should be, which we define as the correct token's logit value being within the top ten logit values at a layer. This experiment aims to answer the overall question by obtaining answers to three related sub-questions: (1) \textit{At what layer does the correct token's logit value first break into the top 10 logits?} (2) \textit{When does the Code LM first consider that the correct token should be predicted as the next token (i.e., the correct token is associated with the highest logit value)?} and (3) \textit{When does the correct token's logit value consistently rank as the highest logit value for all subsequent layers?} {The respective answers to these questions for each of the sub-tasks can be found in Table~\ref{tab:subtask_top_ten_logits}, where for each sub-task we report the median layer (zero-indexed) across all prompts of that sub-task. The logit values in each layer are calculated by applying the logit lens to the residual-stream activation of that layer. We find that the Two Closing Parentheses sub-task has a lower median layer across the considered metrics for all sub-tasks. This was especially apparent for the median first layer where the correct token has the highest overall logit ($L_{\text{Top1}}$) and the median first layer where the correct token is consistently ranked as the top token for all subsequent layers ($L_{\text{ConsistentTop1}}$), where the Two Closing Parentheses sub-task reaches these milestones in layer 25 and the Three Closing Parentheses and Four Closing Parentheses sub-tasks reach these milestones in the final layers of the Code LM. We conjecture that the Code LM views the Two Closing Parentheses sub-task as being easier than the Three and Four Closing Parentheses sub-tasks, which the Code LM appears to view as having similar difficulty.}

\subsection{RQ2: How does each (sub-)layer contribute to the correct token prediction?}\label{subsec:exp2}



\begin{figure*}[t!]
    \centering
    \begin{subfigure}[b]{0.48\textwidth}
        \centering
        \includegraphics[width=\linewidth]{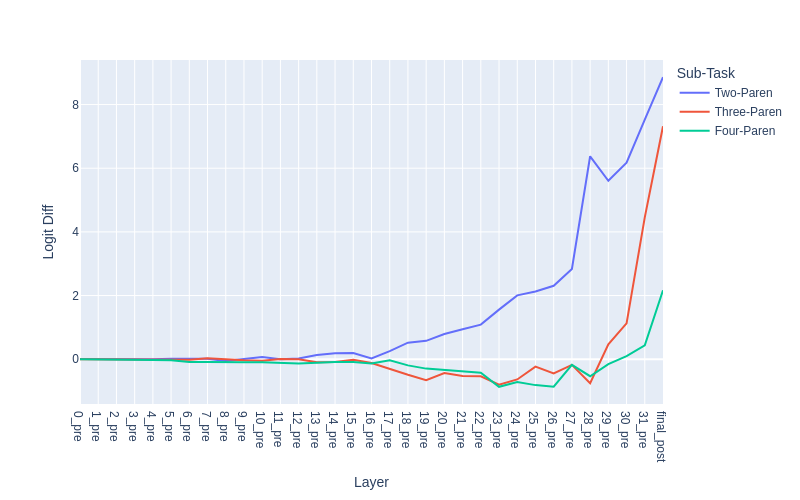} 
        \caption{Logit difference for each sub-task (averaged over prompts of the same sub-task). }
        \label{fig:accum_subtask}
    \end{subfigure}
    \hfill
    \begin{subfigure}[b]{0.48\textwidth}
        \centering
        \includegraphics[width=\linewidth]{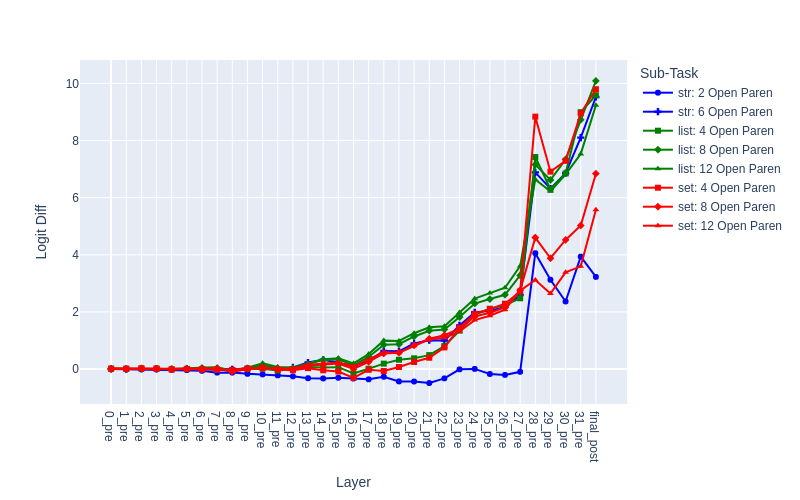} 
        \caption{Logit difference for the Two Closing Paren sub-task, grouped by different prompt types (averaged over prompts of the same type).}
        \label{fig:accum_subtask2}
    \end{subfigure}
    \caption{Logit difference of the Code LM between the correct and the counterfactual tokens across layers of the residual stream. ``$L$\_pre'' and ``$L$\_post'' indicate residual-stream activations before and after layer $L$, respectively.}
    \label{fig:accum_subtask_main}
\end{figure*}

\begin{figure*}[t!]
    \centering
    \begin{subfigure}[b]{0.48\textwidth}
        \centering
        \includegraphics[width=\linewidth]{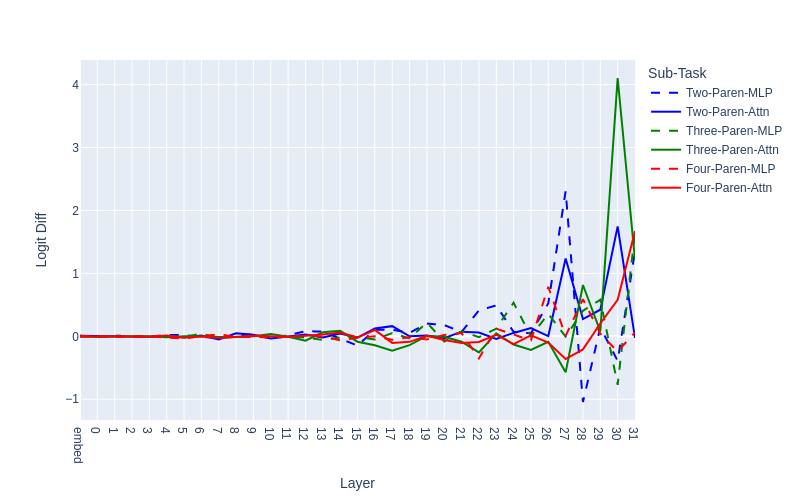} 
        \caption{Sub-layer logit difference for each sub-task (averaged over prompts of the same sub-task). }
        \label{fig:perlayer_subtask}
    \end{subfigure}
    \hfill
    \begin{subfigure}[b]{0.48\textwidth}
        \centering
        \includegraphics[width=\linewidth]{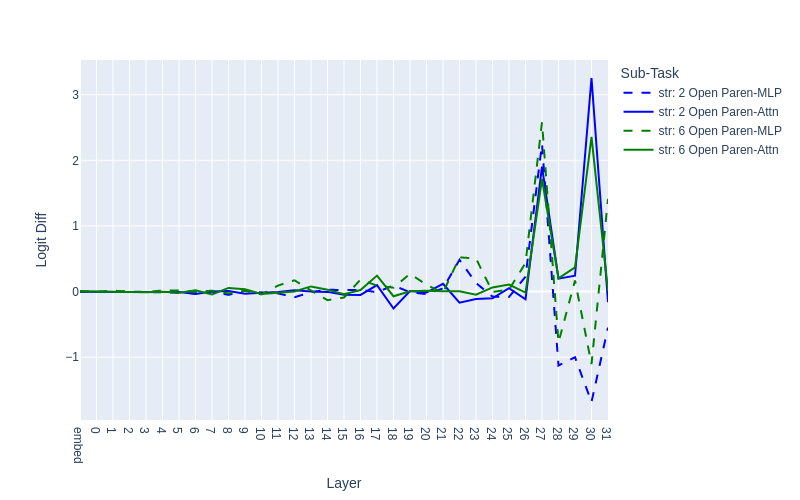} 
        \caption{Sub-layer logit difference for the Two Closing Paren sub-task when the class constructor is \lstinline|str| (averaged over prompts of the same type).}
        \label{fig:perlayer_subtask2}
    \end{subfigure}
    \caption{Sub-layer logit difference of the Code LM between the correct and counterfactual tokens contribution to the residual stream. Figures of sub-layer logit difference for other class constructors are shown in the Appendix~\ref{app: app1}. 
    }
    \label{fig:perlayer_subtask_main}
\end{figure*}

\begin{figure*}[t!]
    \centering
    \begin{subfigure}[b]{0.33\textwidth}
        \centering
        \includegraphics[width=\linewidth]{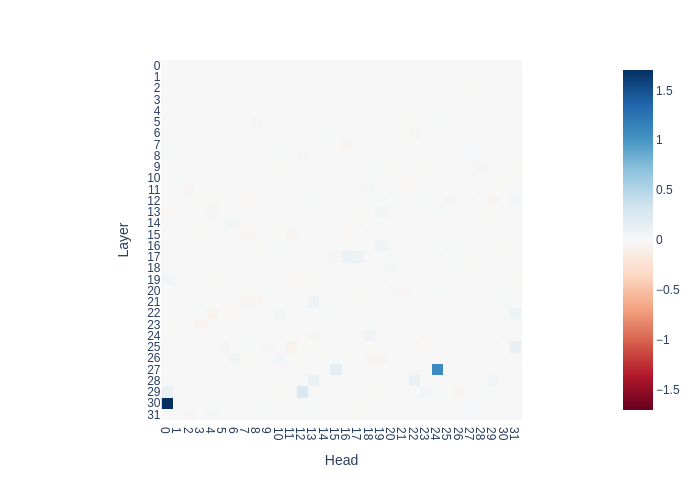} 
        \caption{Two Closing Paren}
        \label{fig:perhead_heatmap_two}
    \end{subfigure}
    \hfill
    \begin{subfigure}[b]{0.33\textwidth}
        \centering
        \includegraphics[width=\linewidth]{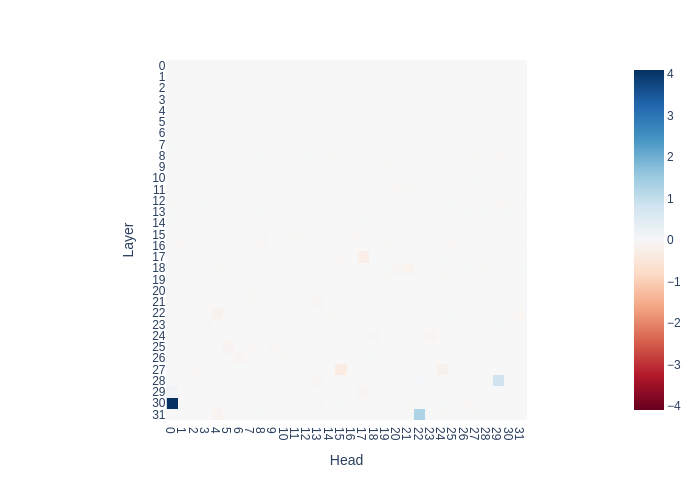} 
        \caption{Three Closing Paren}
        \label{fig:perhead_heatmap_three}
    \end{subfigure}
    \hfill
    \begin{subfigure}[b]{0.33\textwidth}
        \centering
        \includegraphics[width=\linewidth]{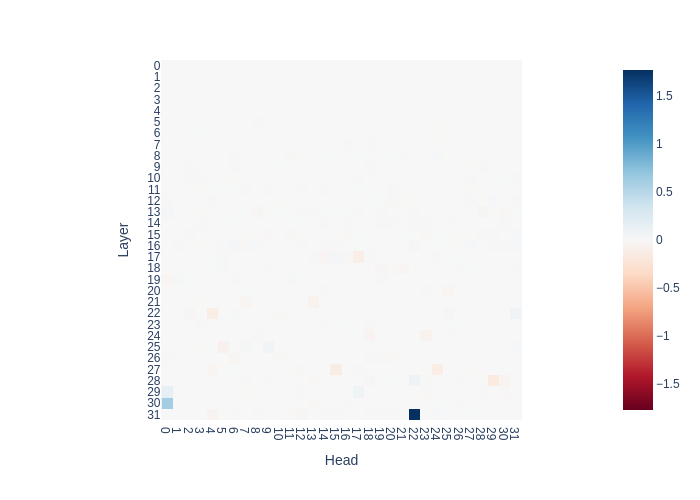} 
        \caption{Four Closing Paren}
        \label{fig:perhead_heatmap_four}
    \end{subfigure}
    \caption{Logit differences between the correct and counterfactual tokens of various attention layers and heads for each sub-task. We observed that the contribution to the logit difference was dominantly made by a few heads (e.g., $L30H0$ and $L27H24$ for the Two Closing Parenthesis task).}
    \label{fig:perhead_heatmap}
\end{figure*}

\begin{table*}[t!]
    \centering
    \resizebox{0.7\textwidth}{!}{
    \begin{tabular}{ccc}
         \toprule
         \textbf{Sub-Task} & \textbf{$L30H0$} & \textbf{$L27H24$}   \\
         \midrule 
          Two Closing Paren & 
            \includegraphics[width=0.25\textwidth]{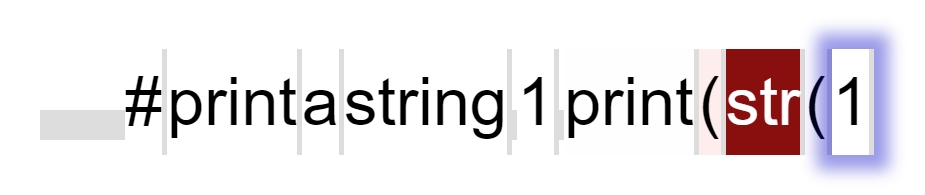} &  
            \includegraphics[width=0.25\textwidth]{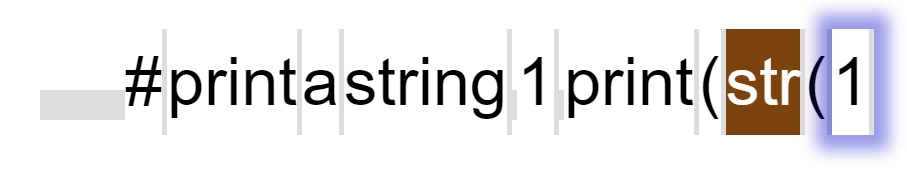}
        \\
         \midrule
         Three Closing Paren & 
            \includegraphics[width=0.33\textwidth]{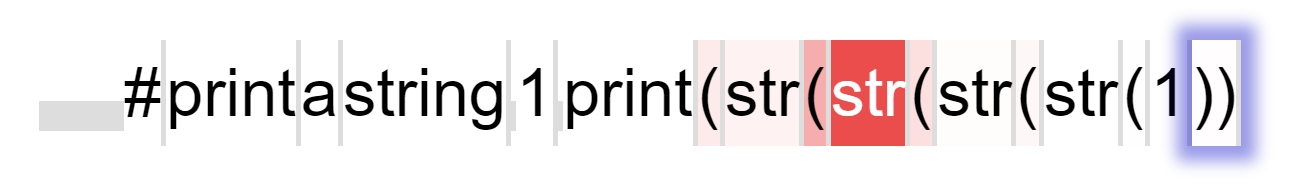} &  
            \includegraphics[width=0.33\textwidth]{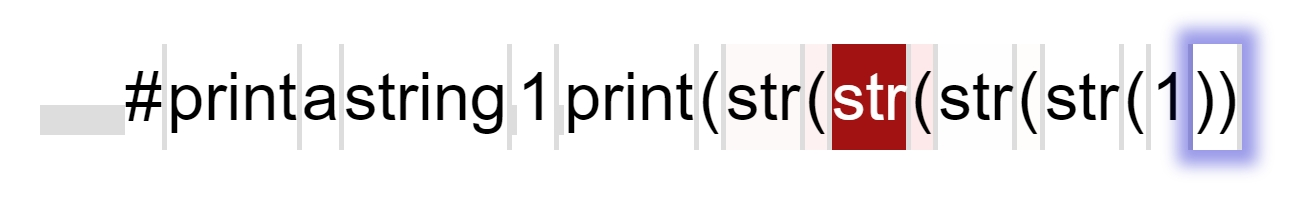}
        \\
         \midrule
         Four Closing Paren & 
            \includegraphics[width=0.3\textwidth]{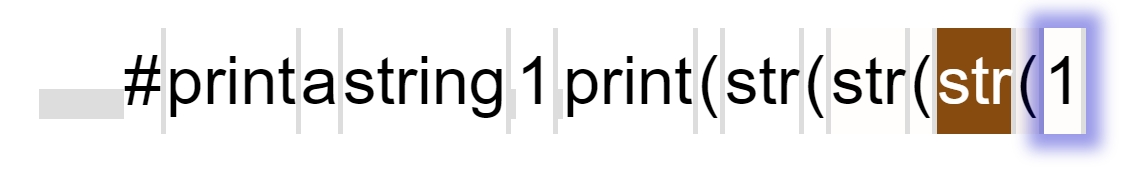} &  
            \includegraphics[width=0.3\textwidth]{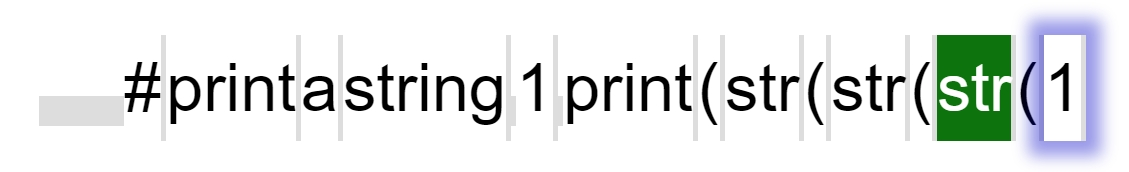}
        \\
         \bottomrule
    \end{tabular}
    }
    \caption{Attention patterns of the attention heads $L30H0$ and $L27H24$ for example prompts from each sub-task. When predicting the next token, both $L30H0$ and $L27H24$ predominantly attend to the innermost unclosed function call across all sub-tasks, suggesting that these attention heads are capable of tracking the number of unclosed parenthesis or function calls.}
    
    \label{tab:attention_patterns}
\end{table*}

\paragraph{Layer-level Analysis}
{To understand each layer's contribution to correct token prediction, we first measure the \emph{logit difference} between the correct token and the counterfactual token on the residual stream across all layers}. Specifically, Figure~\ref{fig:accum_subtask} showcases the logit difference across layers of the residual stream for each sub-task, averaged over prompts in the same sub-task. Figure~\ref{fig:accum_subtask2} displays the same metric but focuses on the results for the Two Closing Parentheses sub-task aggregated at a \emph{prompt type} level---here, we define each prompt type by the class constructor (i.e., \lstinline|str|, \lstinline|list|, or \lstinline|set|) and the number of open parentheses. {The results in Figure~\ref{fig:accum_subtask} show that the Code LM gains an understanding of the Two Closing Parentheses sub-task much earlier than the Three and Four Closing Parentheses sub-tasks, as evidenced by the residual stream having a positive logit difference in earlier layers. The result also corroborates our conjecture from Section~\ref{subsec:exp1} that the Code LM has more difficulty performing the Three and the Four Closing Parentheses sub-tasks. The layers with the largest contributions (defined by ``peaks'' of positive logit difference) to the residual stream for the Two, Three, and Four Closing Parentheses sub-tasks were layers 27 and 30, layers 30 and 31, and layers 31 and 26, respectively. 

The prompt type level results for the Two Closing Parentheses sub-task found in Figure~\ref{fig:accum_subtask2} confirm the positive logit difference contributions of layers 27 and 30 for all prompt types.} {For prompt types that utilize either the \lstinline|list| or \lstinline|set| class constructors, we find that the Code LM gains an understanding of the Two Closing Parentheses sub-task around layer 18. Similar behavior can be seen for the prompt types that utilize the \lstinline|str| class constructor as the number of opening parentheses becomes large, in this case having 6 open parentheses. It appears that as the prompt type becomes sufficiently difficult, whether that be through the inclusion of more open parentheses or utilization of the \lstinline|list| or \lstinline|set| constructor calls, the mid-late layers play a larger role in the promotion of the correct token.}

\paragraph{Sub-Layer Analysis}
The logit difference contribution to the residual stream at a sub-layer (i.e., FF or MHA) level can be seen in Figure~\ref{fig:perlayer_subtask} and Figure~\ref{fig:perlayer_subtask2}. Similar to above, Figure~\ref{fig:perlayer_subtask} displays each of the sub-layer contributions to the residual stream aggregated at a sub-task level, while Figure~\ref{fig:perlayer_subtask2} showcases the results for the Two Parentheses Sub-task when the class constructor is \lstinline|str|.
At the sub-layer level, the logits of each sub-layer were calculated by projecting the sub-layer's immediate activation output to the vocabulary space via the logit lens.
The results in Figure~\ref{fig:perlayer_subtask} reveal the importance of both FF and MHA sub-layers for the closing parentheses task, as all sub-tasks have both FF and MHA sub-layers in their top five positively contributing sub-layers. When comparing the overall contribution of the MHA sub-layers and the FF sub-layers for all sub-tasks, \emph{the MHA sub-layers have a similar or larger positive contribution}---in fact, the most salient negative contributions to the logit difference for the Two and Three Closing Paren sub-tasks both come from MLP. Their positive contribution is especially apparent in the Three Closing Parentheses and Four Closing Parentheses sub-tasks. The MHA sub-layers that exhibit the strongest promotion of the correct tokens against the counterfactual ones for the Two Closing Parentheses, Three Closing Parentheses, and Four Closing Parentheses sub-tasks are the MHA sub-layers in layers 30 and 27, layers 30 and 31, and layers 31 and 30, respectively.

Figure~\ref{fig:perlayer_subtask2} similarly illustrates the importance of MHA sub-layers to positive logit difference. Interestingly, the patterns when the number of open parentheses is 2 and 6 respectively are pretty similar, with MHA's positive contributions peaking at layers 27 and 30, MLP's positive contributions peaking at layers 27 and 22, and MLP's negative contributions dipping significantly at layers 28 and 30. 
\subsection{RQ 3: How do attention heads contribute to the promotion/suppression of correct tokens?}\label{subsec:exp3}

Given the insight from the experimental results of RQ2 that MHA sub-layers appear to contribute in a more significant fashion to the syntax completion task, we ran an additional experiment to see how individual attention heads contribute to the promotion or suppression of the logit difference between the correct token and the counterfactual token. We were especially interested in identifying the attention heads that had a large positive or negative contribution to the promotion of the logit difference in the previously identified MHA sub-layers. {The heat map showing the logit difference projected from individual attention layers and heads for each sub-task can be seen in Figure~\ref{fig:perhead_heatmap}. {For all sub-tasks, we identify that {most} heads have a strong positive contribution to the logit difference (marked as deep blues), whereas {only a few} heads have a small negative contribution (marked as light reds). In particular, the positive contribution was dominantly made by only a few heads.}
For the Two Closing Parentheses sub-task, we find that the largest contributing heads are $L30H0$ (i.e., layer 30, head 0) and $L27H24$ (i.e., layer 27, head 24), which both have positive contributions. {Interestingly, while the $L30H0$ attention head exhibits similar positive contribution behavior in the Three and Four Closing Parentheses sub-tasks as in the Two Closing Parentheses sub-task, we find that the $L27H24$ head negatively contributes the correct output for Three Closing Parentheses and Four Closing Parentheses sub-tasks.}

{We identified that, despite their functional differences, $L30H0$ and $L27H24$ have similar attention patterns for all sub-tasks. Specifically, they both were found to effectively track the number of already closed parentheses by attending to the function name up to the point where the parentheses are already closed, as shown in Table~\ref{tab:attention_patterns}. However, while $L30H0$ dynamically promotes the correct number of closing parentheses based on the count of those already closed, $L27H24$ always promotes two closing parentheses regardless of the number of remaining open parentheses. In other words, this head is promoting incorrect knowledge despite being able to correctly understand the context. We summarize this the phenomenon as \emph{``incorrect knowledge association''}. Consequently, this behavior of $L27H24$ results in a negative contribution for tasks requiring three or four closing parentheses.}

\section{Related Works}

{\paragraph{Analysis of Code LMs} Code LMs~\cite{abdin2024phi, team2024codegemma, guo2024deepseek, li2023starcoder, roziere2023code, chen2021evaluating}, are a class of LMs specifically developed to enhance code generation capabilities of LMs through fine-tuning and additional training techniques~\cite{chan2023transformer}. Although these models have demonstrated remarkable capabilities in code generation tasks~\cite{yu2024codereval, zhuo2024bigcodebench, lai2023ds, cassano2023multipl, hao2022aixbench, srivastava2022beyond, hendrycks2021measuring}, they remain susceptible to various syntactic and semantic (or logical) errors~\cite{yu2024codereval, tambon2024bugs, dou2024s}. Prior studies have focused on empirically examining various types of bugs across a range of coding tasks and programming languages~\cite{dou2024s, tambon2024bugs, dakhel2023github} or proposing benchmark datasets to characterize these models' shortcomings~\cite{wang2023recode, siddiq2022securityeval, yang2024seccodeplt}. For instance, \citet{dou2024s} observed that LMs are especially prone to syntactical errors (e.g., incomplete syntax structure, and indentation issues) when generating code for complex or lengthy problems. While these studies provide valuable insights into when Code LMs are likely to make mistakes, our understanding of the underlying internal mechanisms enabling code-generation capabilities remains limited. To bridge this gap, our study investigates how Code LMs perform syntax completion tasks.}

{\paragraph{Mechanistic Interpretability (MI)} MI is a subfield of interpretability that aims to reverse-engineer LM by understanding their internal components and computational processes~\cite{elhage2021mathematical, olah2020zoom, rai2024practical, bereska2024mechanistic}. Recent MI studies have investigated various LM behaviors, including sequence completion task~\cite{elhage2021mathematical}, Indirect Object Identification~\cite{wang2022interpretabilitywildcircuitindirect}, Python docstring completion~\cite{heimersheim2023circuit, conmy2023towards} and modular addition tasks~\cite{nanda2023progress}, by discovering circuits, a subset of LM components responsible for implementing these LM behaviors. These circuits can be explained in terms of human-understandable algorithms after interpreting the circuit components, which has led to the discovery of several interpretable attention heads such as induction heads~\cite{elhage2021mathematical}, suppression heads~\cite{mcdougall2023copy}, and previous token heads~\cite{elhage2021mathematical}. Building upon these advancements, we study the internal mechanisms of code LMs to understand the syntax completion capability of Code LMs. As far as we know, there has no been prior work carefully studying MI techniques in the application of LMs for code generation. As the first step, in this work, we have focused on discovering how the CodeLlama model identifies the correct next token and contrasts it against the counterfactual token in the syntax completion task. We include a discussion of our future extension along this line of research in Section~\ref{sec: discussion}.}

\section{Discussion and Future Works} \label{sec: discussion}
{In this work, we presented preliminary findings toward a mechanistic understanding of how Code LMs use their internal knowledge to complete syntactic completion tasks, identifying multiple attention heads that play a critical role in this task. Building on these results, we suggest the following directions for future research.}

{\paragraph{Circuit Discovery}
Seeing the relatively more important role the MHA sub-layers play in prioritizing the correct token over the counterfactual one, our work has focused on analyzing the MHA patterns in a Code LM. However, future work should extend the analysis to cover the MLP sub-layers (which were found to implement knowledge look-up in transformers~\cite{geva2021transformer}), and eventually portray a complete \emph{circuit} of how a Code LM associates various components in its transformer architecture towards successfully using its knowledge in syntax completion.}

{\paragraph{Universality of the Interpretation}
We hypothesize that there is significant overlap among similar sub-tasks involved in syntax completion tasks, both within and across programming languages. For example, we investigated CodeLlama's ability to perform the parenthesis completion task, which requires the model to track the number of open parentheses that have been closed. Similar counting mechanisms might also be needed for managing indentation in Python or closing curly braces in JavaScript. Our future work will look into whether a Code LM reuses the same components across tasks and languages for similar roles.
}

{\paragraph{Improving the LM Performance}
Finally, we aim to utilize our interpretation result to enhance a Code LM's performance in real life.
For example, in experiments we have identified an attention head, $L27H24$, that performs \emph{incorrect knowledge association} and erroneously promotes two closing parentheses even when the model needs to generate three or four closing parentheses. Furthermore, recent work of \citet{geva2022transformer} and \citet{rai2024investigation} have showcased the potential of directly controlling a model's generation or task performance via manipulating its neuron activation. In the future, we will similarly explore if suppressing such less precise attention heads can improve the accuracy of the Code LM in closing the parentheses and beyond.
}

\section*{Acknowledgements}
This project was sponsored by the National Science Foundation (\# 2311468 / \#2423813) and College of Computing and Engineering and the Department of Computer Science at George Mason University. This project was also supported by resources provided by the Office of Research Computing at George Mason University (\url{https://orc.gmu.edu}) and funded in part by grants from the National Science Foundation (2018631). We thank the anonymous reviewers and members at GMU NLP group for their feedback on this work.

\bibliography{main}
\appendix
\section{Additional Results of Sub-Layer Logit Differences} \label{app: app1}

In Figure~\ref{fig:additional-sub-layer}, we present the sub-layer logit difference curves for the other two class constructors, i.e., \lstinline|list| and \lstinline|set|, in the Two Closing Parenthesis sub-task.

\begin{figure*}[t!]
    \centering
    \begin{subfigure}[b]{0.45\textwidth}
        \centering
        \includegraphics[width=\linewidth]{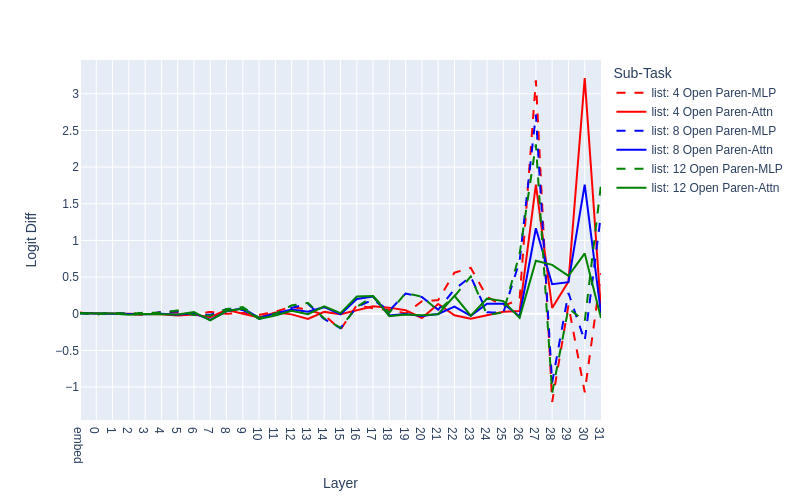} 
        \caption{Sub-layer logit difference for the Two Closing Paren sub-task when the class constructor is \lstinline|list| (averaged over prompts of the same type).}
        \label{fig:list-type}
    \end{subfigure}
    \hfill
    \begin{subfigure}[b]{0.45\textwidth}
        \centering
        \includegraphics[width=\linewidth]{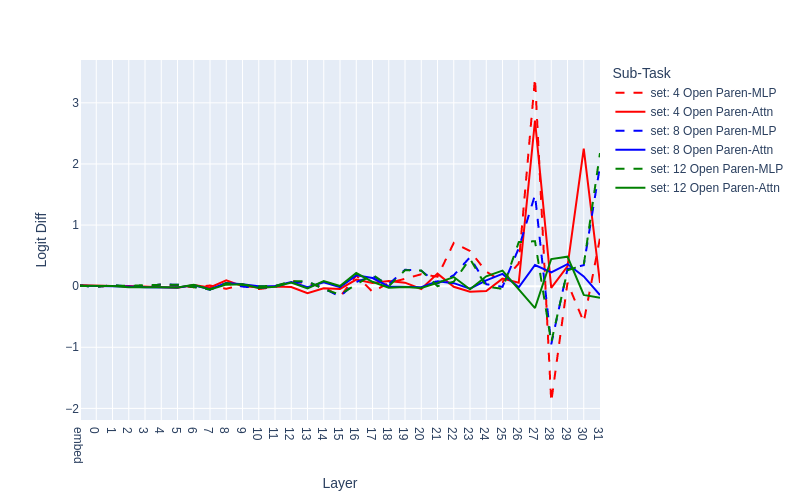} 
        \caption{Sub-layer logit difference for the Two Closing Paren sub-task when the class constructor is \lstinline|set| (averaged over prompts of the same type).}
        \label{fig:set-type}
    \end{subfigure}
    \caption{Sub-layer logit difference of the Code LM between the correct and counterfactual tokens contribution to the residual stream. ``embed'' indicates the word embedding.}
    \label{fig:additional-sub-layer}
\end{figure*}
\end{document}